\documentclass[11pt,prd,superscriptaddress,showpacs,nofootinbib]{revtex4-1}
\usepackage{amsmath}
\usepackage{amsfonts}
\usepackage{versions}
\usepackage[usenames, dvipsnames]{color}
\usepackage{amssymb}
\usepackage{graphicx}
\usepackage{natbib}
\usepackage{hyperref}
\usepackage{lineno}%
\usepackage{color}

\begin{document}

\title{On effective loop quantum geometry of Schwarzschild interior }

\date{\today}

\author{Jer\'onimo Cortez}
\email[E-mail: ]{jacq@ciencias.unam.mx}
\affiliation{Departamento de F\'isica, Facultad de Ciencias, Universidad Nacional Aut\'onoma de M\'exico, Ciudad de M\'exico 04510, M\'exico}

\author{William Cuervo}
\email[E-mail: ]{wfcuervo@gmail.com}
\affiliation{Departamento de F\'isica, Facultad de Ciencias, Universidad Nacional Aut\'onoma de M\'exico, Ciudad de M\'exico 04510, M\'exico}
\affiliation{Departamento de F\'{\i}sica, Universidad Aut\'onoma Metropolitana Iztapalapa,
San Rafael Atlixco 186, CP 09340, Ciudad de M\'exico, M\'exico.}

\author{Hugo A. Morales-T\'ecotl}
\email[E-mail: ]{hugo@xanum.uam.mx}
\affiliation{Departamento de F\'{\i}sica, Universidad Aut\'onoma Metropolitana Iztapalapa,
San Rafael Atlixco 186, CP 09340, Ciudad de M\'exico, M\'exico.}

\author{Juan C. Ruelas}
\email[E-mail: ]{j.carlos.ruelas.v@gmail.com}
\affiliation{Departamento de F\'{\i}sica, Universidad Aut\'onoma Metropolitana Iztapalapa,
San Rafael Atlixco 186, CP 09340, Ciudad de M\'exico, M\'exico.}

\begin{abstract}

The success of loop quantum cosmology  to resolve  classical singularities of
homogeneous models has led to its application to 
classical Schwarszchild black hole interior which takes the form of a homogeneous, Kantowski-Sachs, model. First steps were done in pure quantum mechanical terms hinting at the traversable character of the would be classical singularity and then others were performed using effective heuristic models capturing quantum effects that allowed a geometrical description closer to the classical one but avoiding its singularity. However, the problem to establish the link between the quantum and effective descriptions was left open. 
In this work we propose to fill in this gap by considering the path integral approach to the loop quantization of the Kantowski-Sachs model corresponding to the Schwarzschild black hole interior. We show the transition amplitude can be expressed as a path integration over the imaginary exponential of an effective action which just coincides, under some simplifying assumptions, with the heuristic one.
Additionally we further explore the consequences of the effective dynamics. We prove first such dynamics imply some rather simple bounds for phase space variables  and in turn, remarkably, in an analytical way, they imply
various phase space functions that were singular in the classical model
are now well behaved. In particular, the expansion rate, its time derivative, and shear
become bounded and hence the Raychauduri equation is finite term by term thus
resolving the singularities of classical geodesic congruences. Moreover,
all effective scalar polynomial invariants turn out to be bounded.
\end{abstract}

\pacs{04.60.Pp, 03.65.Sq, 04.70.Dy, 04.70.Bw, 98.80.Qc}

\maketitle

%\classification{PACS numbers: 04.60.Pp, 03.65.Sq, 04.70.Dy, 98.80.Qc}

%%%%%%%%%%%%%%%%%%%%%%%%%%%%%%%%%%%%%%%%%%%%
%% MAINMATTER
%%%%%%%%%%%%%%%%%%%%%%%%%%%%%%%%%%%%%%%%%%%%

%\tableofcontents

\pagebreak 

\section{Introduction}

Two main questions of theoretical physics requiring the knowledge of the structure of spacetime at a fundamental level are the nature of singularities appearing in classical general relativity and the ultraviolet divergences of field theory. It is expected a quantum theory of gravity can provide an answer for such questions as we have learned from simpler quantum physical systems which  improve their behavior as compared to their classical analogues. Not only is compulsory to find how these questions can be answered but in fact to be able to grasp what new concepts, if any, are needed in the theoretical framework that replaces the origin of these issues.

One candidate quantum gravity theory, loop quantum gravity (LQG) \cite{Ashtekar-etal(2004), Rovelli200712,Thiemann200812} which is a non-perturbative, background-independent approach to quantize general relativity is natural to consider in dealing with the nature of spacetime.
In particular, the implementation of the loop quantum gravity program for cosmological models, which is known as loop quantum cosmology (LQC) \cite{Bojowald:2006da,GAMM,Ashtekar:2011ni,CorichiAshtekarPetkov}, has led to the replacement of the big-bang singularity with a quantum bounce for homogeneous and isotropic models (see, for instance, the seminal works \cite{Bojowald:2001xe,Ashtekar:2006uz,Ashtekar:2006wn,Ashtekar:2006es,Vandersloot:2006ws}). Also anisotropic  \cite{Chiou:2007sp,Chiou:2007mg,MartinBenito:2008wx, MartinBenito:2009qu, Ashtekar:2009vc, Ashtekar:2009um,WilsonEwing:2010rh,Singh:2011gp, Fujio:2012zz, Liu:2012xp,Corichi:2012hy, Corichi:2015ala,Singh:2013ava,Modesto:2005zm,Ashtekar:2005qt,Bohmer:2007wi,Campiglia:2007pb,Campiglia:2007pr,Chiou:2008eg,Chiou:2008nm,Modesto:2008im,Gambini:2008dy,Cortez:2012ina,Gambini:2013hna,Gambini:2013ooa,Joe:2014tca,Corichi:2015xia}, as well as inhomogeneous models \cite{Bojowald:2006qu,MenaMarugan:2009dp,Garay:2010sk,MartinBenito:2010up,MartinBenito:2010dz,Olmedo:2011zz,Brizuela:2011ps,MartindeBlas:2012zz,Martin-Benito:2013jqa,Tarrio:2013ija,Fernandez-Mendez:2014raa,Fernandez-Mendez:2014aea,Gomar:2014faa,MartIN-Benito:2015pca} have been studied. 

It has been argued in some cases \cite{Ashtekar:2011ni} it is convenient to use effective models capturing their essential quantum aspects mainly when the full quantum dynamics is unknown. The effective approach has been tested by applying the effective dynamics to cases where quantum evolution is fully known, with the astonishing result that the effective dynamics matches quite well, even in the deep quantum regime, with the full quantum dynamics of LQC \cite{Ashtekar:2006wn,Bentivegna:2008bg,Ashtekar:2006es,Diener:2014mia}. Clearly it is crucial to determine whether and when an effective description is pertinent without relying on the full quantum solution.

Motivated by the success of LQC in the study of homogeneous cosmologies, the study of the Schwarzschild black hole interior by using LQC techniques was put forward in  \cite{Modesto:2005zm,Ashtekar:2005qt} exploiting the fact that the interior Schwarzschild geometry is a particular homogneous Kantowski-Sachs model. Their results indicated that quantum Einstein equations were not singular. However, the answer to the question what replaces the classical singularity was not answered. Further developments using an effective approach were done in \cite{Bohmer:2007wi,Campiglia:2007pb,Campiglia:2007pr,Chiou:2008eg,Chiou:2008nm,Modesto:2008im,Gambini:2008dy,Cortez:2012ina,Gambini:2013hna,Gambini:2013ooa,Joe:2014tca,Corichi:2015xia} (For a recent review see \cite{Olmedo:2016ddn}) . Interestingly, \cite{Bohmer:2007wi} argued that there is a connection between the black hole and a Nariai Universe whereas \cite{Corichi:2015xia} found the presence of a white hole instead; the difference between these two works being whether a pair of parameters in the quantization are scale factor dependent or constants, respectively. In particular \cite{Bohmer:2007wi} get quantum corrections both at singularity and horizon, as opposed to \cite{Corichi:2015xia} which corrections are limited to the would be singularity. Actually such difference appears already in cosmological models in regard to the inadequacy of the so called  $\mu_{0}=constant$ prescription \cite{Ashtekar:2011ni,CorichiAshtekarPetkov}; in order to correctly describe the classical regime such parameter should be scale factor dependent.  Yet  \cite{Corichi:2015xia} argue such analysis for cosmological models does not hold for Schwarzschild since it would alter the notion of classical horizon. To us this is an unsettled issue which requires further study and more information, for instance,  the  behavior of effective quantities, like geometric scalars in the effective Raychaudhuri equation, or the effective Kretschmann and curvature scalars (see \cite{Cortez:2012ina,Joe:2014tca} for first steps in this direction). 

The present work is aimed at filling the gap between the description using loop quantum model and that using an effective dynamics for the Kantowski-Sachs model representing the Schwarzschild black hole interior. We will derive the effective Hamiltonian constraint via the path integral approach starting from the quantum Hamiltonian in the so called improved dynamics scheme with the quantization parameters depending on the scale factors and consider the transition amplitude between two basis states labelled with different values of a time parameter. After performing the usual partition of the time interval we get the effective action, $S_{{\rm{eff}}}$, as the argument of an imaginary exponential that is to be integrated upon according to Feynman's prescription. It is from $S_{\rm{eff}}$ that the effective Hamiltonian $H_{\rm{eff}}$ will be extracted. Thereafter we will analyze in an analytical manner the effective Hamiltonian theory associated to $H_{\rm{eff}}$ and its impact on the behavior of relevant scalars. More precisely, we will prove that the effective expansion scalar, its time derivative and shear  are bounded. Moreover, it is demonstrated that every scalar polynomial invariant, so, in particular, the Ricci and Kretschmann scalars, are bounded in the effective approach. 

 This paper is organized as follows. Section \ref{ClassicalKS} is devoted to the classical setting of the theory. We start by recalling that the Schwarzschild interior geometry can be described by a Kantowski-Sachs model. Thereafter, we recast the model in connection variables and perform a qualitative canonical analysis of the classical dynamics identifying the singular behavior of curvature invariants. Next, in Section \ref{QKS}, within the framework of the improved dynamics prescription, we get the effective Hamiltonian constraint by using the path integral approach. This along the lines of \cite{Liu:2012xp} for Bianchi I. The effective loop quantum black hole interior geometry is analyzed in Section \ref{EffectiveKS}, where it is shown that classically divergent quantities are actually bounded in the effective approach. In Section \ref{sec-disc} we discuss and summarize our main results. 
 
 Throughout this work we will denote by $\mu$ to the so-called improved dynamics for homogeneous models, which has previously been denoted in literature by $\bar{\mu}^{\prime}$ (see, for instance, \cite{Chiou:2008nm}).

\section{Classical Theory}

\label{ClassicalKS}

\subsection{The interior geometry in connection variables}

As it is well known, a Schwarzschild black hole of mass $M$ (i.e., a spherically symmetric vacuum solution to general relativity) is described by the metric
\begin{equation}
ds^{2}=-\left(  1-\frac{2GM}{r}\right)  dT^{2}+{\left(  1-\frac
{2GM}{r}\right)^{-1}  }dr^{2}+r^{2}\left(d\theta^{2}+\sin^{2}\theta d\phi^{2}\right), \label{metrica schwarschild}
\end{equation}
in Schwarzschild coordinates $T\in {\mathbb{R}}$, $r\in \mathbb{R}^{+}$, $0\leq \theta \leq \pi$ and $0\leq \phi \leq 2\pi$. At $r=0$ there is a true singularity (the Kretschmann scalar blows up as $r\to 0$) which is wrapped by an event horizon located at the so-called Schwarzschild radius, $r_{s}=2GM$ (where the Schwarzschild coordinates become singular). The exterior (i.e., $r>r_{s}$) spacelike $(\partial / \partial r)^{a}$ and timelike $(\partial / \partial t)^{a}$ vector fields switch into, respectively, timelike and spacelike vector fields at the black hole interior (i.e., $0<r<r_{s}$). For the sake of clarity, let us then rename the interior spatial coordinate $T$ by $x$, and the interior time coordinate $r$ by $t$. Thus, the Schwarzschild interior metric can be written as follows 
\begin{equation}
ds^{2}=-\left(  \frac{2GM}{t}-1\right)^{-1}dt^{2}+\left(  \frac{2GM}
{t}-1\right)  dx^{2}+t^{2}\left(d\theta^{2}+\sin^{2}\theta d\phi^{2}\right),\label{schwarzschild interior}
\end{equation}
where $0<t<2GM$ and $x\in \mathbb{R}$. The singularity now corresponds to an {\emph{initial}} singularity at time $t=0$, resembling to a cosmological singularity. In fact, the Schwarzschild interior solution (\ref{schwarzschild interior}) belongs to the class of Kantowski-Sachs cosmological models \cite{Kantowski:1966te}
with homogeneous spatial sections $\Sigma\approx\mathbb{R}\times S^{2}$, i.e., Kantowski-Sachs models with symmetry group $\mathbb{R}\times SO(3)$, which are described by metrics of the form 
\begin{equation}
ds^{2}=-N^{2}dt^{2}+X^{2}dx^{2}+Y^{2}\left(d\theta^{2}+\sin^{2}\theta d\phi^{2}\right).\label{KS general metric}
\end{equation}
Here, the metric coefficients $X$, $Y$ and the lapse function $N$, depend on the coordinate time $t$ only. 

Since the Schwarzschild interior geometry can be understood as a Kantowski-Sachs spacetime with symmetry group $\mathbb{R}\times SO(3)$, let us consider the Kantowski-Sachs symmetry reduction of canonical general relativity in connection variables to get the connection description for the black hole interior. Following the procedure for the study of homogeneous models \cite{ashbolew} let us introduce an auxiliary metric $\mathring{q}_{ab}$ on the $3$-manifold $\mathbb{R}\times S^{2}$, with compatible triad $\mathring{e}^{a}_{i} $ and co-triad $\mathring{\omega}^{i}_{a}$ that are left-invariant under the action of the Killing fields of $\Sigma$. They carry the symmetry information but ignore the particularities about the minusuperspace model. The usual choice for the auxiliary metric{\footnote{For a different choice see \cite{Corichi:2015xia}.}}, which we will also consider here, is \cite{Ashtekar:2005qt,Modesto:2005zm,Bohmer:2007wi,Chiou:2008eg}:
\begin{equation}
\label{fid-metric}
\mathring{q}_{ab}dy^{a}dy^{b}=dx^{2}+d\theta^{2}+\sin^{2}\theta d\phi^{2}.
\end{equation}
The determinant of the fiducial metric (\ref{fid-metric}) is given by $\mathring{q}=\sin^{2}\theta$, so that the densitized triad $\mathring{E}^{a}_{i}=\sqrt{\mathring{q}}\: \mathring{e}^{a}_{i}$ reads              
$\mathring{E}^{a}_{i}=\sin\theta  \mathring{e}^{a}_{i}$. The compatible densitized triad $\mathring{E}^{a}\partial_{a}=\mathring{E}^{a}_{i}\tau^{i}\partial_{a}$, which takes values in the dual of $su(2)$, and its corresponding $su(2)$-valued co-triad $\mathring{\omega}_{a}dy^{a}=\mathring{\omega} ^{i}_{a}\tau_{i}dy^{a}$, are explicitly given by
\begin{equation}
\label{fiducial-cotriad-triad}
\mathring{\omega}_{a}dy^{a}=\tau_{3}dx + \tau_{2} d\theta - \tau_{1}\sin \theta d\phi,\quad 
\mathring{E}^{a}\partial_{a}=\tau_{3}\sin\theta\partial_{x}+\tau_{2}\sin\theta\partial_{\theta}-\tau_{1}\partial_{\phi},
\end{equation}
where $\tau_i$ are the standard generators of $SU(2)$, satisfying $[\tau_{i},\tau_{j}]=\epsilon_{ij}^{\:\:\: k}\tau_{k}$. 

Note that integrals over $\mathbb{R}\times S^{2}$ involving spatially homogeneous quantities will generally diverge, given the non-compact character of the $x$-direction. To circumvent this feature, which, for instance, is an obstacle to properly calculate the Poisson brackets, one restricts $x$ to an interval of finite length $L$, w.r.t. the fiducial metric, and then perform all integrations over a finite-sized cell ${\cal{V}}_{0}=[0,L]\times S^{2}$ of fiducial volume $V_{0}=4\pi L$. 

Now, by imposing the Kantowski-Sachs symmetry group $\mathbb{R}\times SO(3)$ in the full theory, one gets that the symmetric connection $A=A^{i}_{a}\tau_{i}dy^{a}$ and triad $E=E_{i}^{a}\tau^{i}\partial_{a}$ can be written, after gauge fixing of the Gauss constraint, as follows
\begin{equation}
\label{AE-in-fidu}
A =L^{-1}\,c \left( \mathring{\omega}_{x}dx\right)+b\left( \mathring{\omega}_{\theta}d\theta + \mathring{\omega}_{\phi}d\phi\right)+\Gamma, \quad 
E=p_{c}\left(\mathring{E}^{x}\partial_{x}\right)+L^{-1}p_{b}\left(\mathring{E}^{\theta}\partial_{\theta}-\mathring{E}^{\phi}\partial_{\phi}\right).
\end{equation}
Here, $\Gamma=\Gamma^{i}_{a}\tau_{i}dy^{a}=\cos\theta\tau_{3}d\phi$ is the spin-connection compatible with the triad density $E$. Coefficients $b$, $c$, $p_{b}$ and $p_{c}$, which are all only functions of time, capture the non-trivial information about the symmetry reduced model. From Eqs.(\ref{fiducial-cotriad-triad})-(\ref{AE-in-fidu}), it follows that the Kantowsi-Sachs connection and triad are explicitly given by 
\begin{equation}
A =
L^{-1}\, c\tau_{3}dx+b\tau_{2}d\theta-b\tau_{1}\sin\theta d\phi+\tau_{3}
\cos\theta d\phi, \label{A reducido}
\end{equation}
\begin{equation}
E= p_{c}\tau_{3}\sin\theta\,\partial_{x}+L^{-1}p_{b}\tau_{2}\sin
\theta\,\partial_{\theta}-L^{-1}p_{b}\tau_{1}\,\partial_{\phi}.
\label{E reducido}
\end{equation}

The phase space resulting from the symmetry reduction and gauge fixing processes is  the symplectic space ${\mathbf{\Gamma}}=[(b,p_{b},c,p_{c}),\Omega]$, with symplectic form \cite{Ashtekar:2005qt,Chiou:2008eg}
 \begin{equation}
 \label{reduced-symps}
 \Omega=\frac{1}{8\pi G\gamma}\int_{{\cal{V}}_0}d^{3}y\: \left(dA^{i}_{a}\wedge dE^{a}_{i}\right)=\frac{1}{2G\gamma}\left(dc\wedge dp_{c}+2db\wedge dp_{b} \right),
 \end{equation}
where $\gamma$ is the so-called Barbero-Immirzi parameter. The only non vanishing Poisson brackets defined by the reduced symplectic form (\ref{reduced-symps}) are
\begin{equation}
\label{poisson-brackets}
\{b,p_{b}\}=G\gamma, \quad
\{c,p_{c}\}=2G\gamma.
\end{equation}

Let us remark that, in fact, we will not consider the whole of the phase space $\bf{\Gamma}$. Indeed, as a part of the gauge-fixing procedure, $p_b$ can be chosen to be a strictly positive function \cite{Chiou:2008nm}, $p_{b}>0$. Besides, since distinct signs of $p_c$ correspond to regions with triads of opposite orientations \cite{Ashtekar:2005qt,Chiou:2008nm}, then $p_{c}$ can be chosen to be strictly positive as well. 

Recall that a $3$-metric $q_{ab}$ is related with its compatible densitized triad $E^{a}_{i}$ by $qq^{ab}=E^{a}_{i}E^{b}_{i}$. Thus, from Eqs. (\ref{KS general metric})  and (\ref{E reducido}) it follows that $X^{2}=p_{b}^{2}/(L^{2}p_{c})$ and $Y^{2}=p_{c}$; i.e., in terms of the triad variables, the Kantowski-Sachs metric reads 
\begin{equation}
ds^{2}=-N(t)^{2}dt^{2}+\frac{p_{b}^{2}}{L^{2}p_{c}}dx^{2}+p_{c}\left(
d\theta^{2}+\sin^{2}\theta d\phi^{2}\right)  . \label{metrica cinematica}
\end{equation}
Thus, with respect to the metric (\ref{metrica cinematica}), the length of the interval $[0,L]$ in the $x$-direction, the area of $S^{2}$ and the volume of the cell ${\cal{V}}=[0,L]\times S^{2}$, are respectively given by
\begin{equation}
\label{leng-are-vol}
l=p_{b}/\sqrt{p_{c}},\quad A_{S^{2}}=4\pi p_{c} , \quad V=4\pi p_{b}\sqrt{p_{c}}
\end{equation}

Now, in terms of the reduced canonical variables, $(b,p_{b})$ and $(c,p_{c})$, the Hamiltonian constraint takes the form \cite{Ashtekar:2005qt}
\begin{equation}
{\cal{C}}_{\rm{Ham}}=16\pi G\,H_{{\rm{class}}}=-\frac{8\pi N}{\gamma^{2}}\left[  2bc\sqrt{p_{c}}+(b^{2}+\gamma^{2}
)\frac{p_{b}}{\sqrt{p_{c}}}\right] , \label{class-H}
\end{equation}
which defines $H_{{\rm{class}}}$. By choosing the lapse function equal to one, from Eqs.(\ref{poisson-brackets}) and (\ref{class-H}) we obtain that the dynamics is dictated by
\begin{eqnarray}
\dot{b}  &  = & \{b,H_{{\rm{class}}}\} = -\frac{1}{2\gamma\sqrt{p_{c}}}\:\left(b^{2}+\gamma^{2}\right),\label{bdot} \\
\dot{c}  &  =& \{c,H_{{\rm{class}}}\} =\frac{1}{2\, \gamma\,p_{c}^{3/2}}\:\left(b^{2}p_{b}-2bcp_{c}+\gamma^{2}p_{b}\right),\label{cdot}\\
\dot{p}_{b}  &  =& \{p_{b},H_{{\rm{class}}}\} =\frac{1}{\gamma{\sqrt{p_{c}}}}\:\left(bp_{b}+cp_{c}\right),\label{pbdot}\\
\dot{p}_{c}  &  =&\{p_{c},H_{{\rm{class}}}\} =\frac{1}{\gamma}\:\left(2b\sqrt{p_{c}}\right). \label{pcdot}
\end{eqnarray}

A direct calculation shows that the Ricci and Kretschmann scalars of the metric (\ref{metrica cinematica}) are, respectively, 
\begin{equation}
R=\frac{2p_{c}\ddot{p}_{b}+p_{b}\left(  2+\ddot{p}_{c}\right)
}{p_{b}p_{c}}, \label{ricciscalarkinematic}
\end{equation}
\begin{align}
K =R_{abcd}R^{abcd} &  =\frac{1}{2\text{$p_{b}$}^{2}\text{$p_{c}$}^{4}}\left[
4\text{$p_{c}$}^{2}\left(  3\dot{p}_{b}^{2}\dot{p}_{c}^{2}-4\text{$p_{c}$}
\dot{p}_{b}\dot{p}_{c}\text{$\ddot{p}_{b}$}+2\text{$p_{c}$}^{2}\text{$\ddot
{p}_{b}$}^{2}\right)  \right. \label{Kretschmannscalarkinematic}\\
&  +4\text{$p_{b}p_{c}$}\left(  \text{$p_{c}\ddot{p}_{b}$}\left(  3\dot{p}
_{c}^{2}-2\text{$p_{c}$}\ddot{p}_{c}\right)  +\dot{p}_{b}\left(  -4\dot{p}
_{c}^{3}+2\text{$p_{c}$}\dot{p}_{c}\ddot{p}_{c}\right)  \right) \nonumber\\
&  \left.  +\text{$p_{b}$}^{2}\left(  7\dot{p}_{c}^{4}+2\text{$p_{c}$}\dot
{p}_{c}^{2}\left(  2-5\ddot{p}_{c}\right)  +\text{$p_{c}$}^{2}\left(
8+6\ddot{p}_{c}^{2}\right)  \right) \right]. \nonumber
\end{align}
It is not difficult to see, by using Eqs.(\ref{pbdot})-(\ref{pcdot}), that $\ddot{p}_{b} =bc/\gamma^{2}$ and $\ddot{p}_{c}    =(b/\gamma)^{2}-1$. Substituting the latter expressions into Eqs.(\ref{ricciscalarkinematic})-(\ref{Kretschmannscalarkinematic}), as well as by imposing the constraint (\ref{class-H}) and employing the dynamical equations (\ref{bdot})-(\ref{pcdot}), we get that on the constraint surface
\begin{equation}
\label{class-RK}
R=0, \qquad K=\frac{12}{\gamma^{4}}\left( \frac{b^{2}+\gamma^{2}}{p_{c}}\right)^{2}.
\end{equation}
Solving the equations (\ref{bdot}) and (\ref{pcdot}), (see Eq. (\ref{pcb}) below) one gets $(b^{2}+\gamma^{2})^{2}=a_{0}/p_{c}$, with $a_{0}$ being a constant depending on initial conditions and $\gamma$. Hence the Kretschmann scalar goes as $1/p_{c}^{3}$. Explicitly,
\begin{equation}
\label{class-RK-2}
K=\frac{12a_{0}}{\gamma^{4}p_{c}^{3}}.
\end{equation}
Thus, the Kretschmann scalar blows up as $p_{c}$ tends to zero, corresponding to the classical singularity.

Let us now examine the solutions to the system (\ref{class-H})-(\ref{pcdot}), and let us inspect the behavior of the expansion scalar and shear.

\subsection{Solutions, expansion scalar and shear}

\label{Classical Qualitative Analysis}

To start, notice that $cp_{c}$ is a constant on the constraint surface. Indeed, from Eqs.(\ref{cdot}) and (\ref{pcdot}) it follows that
\begin{equation}
\{cp_{c},H_{{\rm{class}}}\}=-\frac{\gamma}{16\pi}{\cal{C}}_{\rm{Ham}}.
\end{equation}
Let us denote the constant $cp_{c}$ by $\gamma K_{c}$; i.e.,
\begin{equation}
\label{Kc}
cp_{c}=\gamma K_{c}.
\end{equation}
Since the sign flipping $K_{c}\to -K_{c}$ is associated to the time reversal $t\to -t$ \cite{Chiou:2008nm}, the two regions, $K_{c}>0$ and $K_{c}<0$, are causally disconnected. Let us consider the region $K_{c}>0$, as in \cite{Chiou:2008nm} (for the sake of completeness, we will also discuss the opposite choice, $K_{c}<0$, at the end of the present section). Provided that $p_{c}> 0$, we have that $c$ must be a strictly positive function of time $t$. Since ${\cal{C}}_{\rm{Ham}}=0$ implies that $b$ and $c$ must have opposite signs (see Eq. (\ref{class-H})), we then conclude that $b<0$. On the other hand, viewed as a quadratic equation in $b$, the constraint (\ref{class-H}) has discriminant 
$D=\gamma^{2}(K^{2}_{c}-p^{2}_{b}),$
where we have used (\ref{Kc}). Thus, to keep $b$ real, $D$ must be non-negative, which implies that $p_{b}$ is bounded from above
\begin{equation}
p_{b}\leq K_{c}. \label{cotapbclasica}
\end{equation}

Now, note that Eqs.(\ref{bdot}) and (\ref{pcdot}) are actually decoupled equations from the rest of Hamilton's equations. Thus, we have that
\begin{equation}
\frac{dp_{c}}{db}=-\frac{4bp_{c}}{\left(b^{2}+\gamma^{2}\right)}, \label{dpc/db}
\end{equation}
which solution is given by 
\begin{equation}
p_{c}=p_{c0}\left(  \frac{b_{0}^{2}+\gamma^{2}}{b^{2}+\gamma^{2}}\right)
^{2}. \label{pcb}
\end{equation}
Here, $p_{c0}$ and $b_{0}$ stand for initial conditions at $t=t_{0}$. Since $\dot{b}<0$ (c.f. Eq.(\ref{bdot})), we have that $b$ is a monotonically decreasing function of time $t$ and, by virtue of Eq.(\ref{pcb}), so is $p_c$. Now, substituting Eq.(\ref{pcb}) in Eq.(\ref{bdot}), we get that
\begin{equation}
\label{b-dot-sol}
\dot{b}=-\alpha_{0}\left(b^{2}+\gamma^{2}\right)^{2}, \qquad \alpha_{0}=\left[2\gamma \sqrt{p_{c0}}\,(b_{0}^{2}+\gamma^{2})\,\right]^{-1}.
\end{equation}
So that,
\begin{equation}
\label{b-sol}
g(b)=-2\gamma^{3}\alpha_{0}(t-t_{0})+g(b_{0}),\qquad g(s)=\frac{\gamma s}{(s^{2}+\gamma^{2})}+\arctan\left(\frac{s}{\gamma}\right).
\end{equation}
Clearly, $g<0$ for all $b<0$. It is easy to see that $g$ decreases monotonically as $b$ evolves in time, which in turn implies that $g$ is a monotonous decreasing function of $t$. Indeed, a straightforward calculation shows that $dg/dt=-2\gamma^{3}\alpha_{0}$; i.e., $g$ is a monotonically decreasing function of time. Note, in addition, that $-\pi /2 < g$. Thus, the relationship (\ref{b-sol}) makes sense (i.e., it is a well-defined relationship providing $b(t)$ at each given $t$ value) only if $2\gamma^{3}\alpha_{0}\Delta t<\pi/2+g(b_{0})$, where $\Delta t =(t-t_{0})$. Hence, as $\Delta t$ approaches to the maximal value $\Delta t_{f}$,
\begin{equation}
\label{time singularity}
\Delta t _{f}=\frac{1}{2\gamma^{3}\alpha_{0}}\left[\frac{\pi}{2}+g(b_{0})\right],
\end{equation}
the solution $b(t)$ will tend to $b\to -\infty$. Then, from Eq.(\ref{pcb}) it follows that  the solution $p_{c}(t)$ will tend to zero as $\Delta t$ approaches $\Delta t _{f}$. By substituting Eq.(\ref{pcb}) into $c=\gamma K_{c}/p_{c}$  [see Eq.(\ref{Kc})], we get that
\begin{equation}
c=\frac{\gamma K_{c}}{p_{c0}}\left(  \frac{b^{2}+\gamma^{2}}{b_{0}^{2}+\gamma^{2}}\right)
^{2}. \label{cb}
\end{equation}
Hence, the solution $c$ must tend to infinity as  $\Delta t \to \Delta t _{f}$. By using Eq.(\ref{Kc}) and Eq.(\ref{pcb}) into the constraint equation ${\cal{C}}_{\rm{Ham}}=0$ [see Eq.(\ref{class-H})], we obtain that 
\begin{equation}
\label{pbb}
p_{b}=-2\gamma K_{c}\frac{b}{(b^{2}+\gamma^{2})}. 
\end{equation}
Thus, in the limit when $\Delta t$ tends to $\Delta t_{f}$, the solution $p_b$ goes to zero as $1/\vert b \vert$. (Note that $p_{b}/p_{c}$ diverges as $\vert b \vert^{3}$ when $\Delta t \to \Delta t_{f}$).

Relations (\ref{pcb}), (\ref{b-sol}), (\ref{cb}) and (\ref{pbb}) provide the solution to the system (\ref{class-H})-(\ref{pcdot}). Once the solution $b(t)$ is obtained from Eq.(\ref{b-sol}), the rest of solution functions, namely $p_{c}(t)$, $c(t)$ and $p_{b}(t)$, are determined by substituting $b(t)$ into equations (\ref{pcb}), (\ref{cb}) and (\ref{pbb}), respectively. The time domain of the solution functions is $t\in [t_{0},t_{0}+\Delta t]$, with $\Delta t \leq \Delta t_{f}$; given an initial data $(b_{0},p_{b0},c_{0},p_{c0})$ at $t=t_0$, with $b_{0}\in \mathbb{R}^{-}$, $p_{b0}\in (0,K_{c})$, and $c_{0},p_{c0}\in \mathbb{R}^{+}$, the solution will tend to the `endpoint' 
\begin{equation}
(b\rightarrow-\infty,p_{b}\to 0 ,c\to \infty,p_{c}\to0), \label{endpt}
\end{equation}
as $t$ approaches $t_{f}=t_{0}+\Delta t_{f}$. From (\ref{leng-are-vol}), (\ref{pcb}) and (\ref{pbb}), it follows that the length $l$, the area $A_{S^{2}}$ and the cell volume $V$ will behave as $l\sim \vert b \vert$, $A_{S^{2}}\sim 1/b^{4}$ and $V\sim 1/\vert b \vert^{3}$ as $t\to t_{f}$.

Let us now consider the congruence of timelike geodesics defined by comoving observers in Kantowski-Sachs spacetime (\ref{metrica cinematica}), with $N=1$; that is, the associated vector field to the congruence is $\xi^{a}=(\partial/\partial t)^{a}$. Thus, 
the expansion scalar $\theta$ corresponds to $\dot{V}/V$, where $V=4\pi p_{b}\sqrt{p_{c}}$ is the congruence's cross-sectional volume. A simple calculation shows that
\begin{equation}
\label{scalar-factor-r}
\theta=\frac{\dot{p}_{b}}{p_{b}}+\frac{\dot{p}_{c}}{2p_{c}}.
\end{equation}
By using Eqs.(\ref{bdot}), (\ref{pcdot}), (\ref{pcb}) and (\ref{pbb}), as well as calculating $\dot{p}_b$ by employing Eq.(\ref{pbb}), it is not difficult to see that
\begin{equation}
\label{class-exp-scalar}
\theta=\alpha_{0}\left(\frac{b^{2}+\gamma^{2}}{b}\right)\left(3b^{2}-\gamma^{2}\right).
\end{equation}
Clearly, the expansion scalar is a monotonically decreasing function of time (recall that $b<0$ and that it is a monotonically decreasing function of $t$). What is more, irrespective of the initial condition $b_0$, $\theta\to -\infty$ as $t$ tends to the maximal value $t_{f}$ (i.e., the volume shrinks to zero as $t\to t_{f}$, invariably). At finite proper time $\Delta t_{f}$, the congruence of timelike geodesics develop a caustic, and the cell volume becomes zero; in fact, the geodesics of the congruence turn out to be inextendible (i.e., incomplete). Note, however, that depending upon the initial condition $b_0$, there would be a stage where the volume, in fact, will enlarge. Indeed, observe that
$\theta$ is strictly positive for $-\gamma/\sqrt{3}<b<0$, it is zero at $b=-\gamma/\sqrt{3}$, and it is strictly negative for $b<-\gamma/\sqrt{3}$. Thus, if the initial condition $b_{0}$ is in $(-\gamma/\sqrt{3},0)$, we will have that the volume will increase up to a maximum value $V_{{\rm{max}}}$, at $b=-\gamma/\sqrt{3}$, and afterwards it will monotonically decrease up to zero volume, at $t=t_{f}$ (time at which $p_{c}$ vanishes and the Kretschmann scalar blows up). 

A direct calculation shows that the time derivative of the expansion scalar (\ref{class-exp-scalar}) is given by 
\begin{equation}
\label{class-theta-dot}
\dot{\theta} =-\alpha_{0}^{2}\left(\frac{b^{2}+\gamma^{2}}{b}\right)^{2} \left(9b^{4}+2\gamma^{2}b^{2}+\gamma^{4}\right),
\end{equation}
where we have used (\ref{b-dot-sol}). Since $\dot{\theta}$ is strictly negative, the expansion scalar $\theta$ is a monotonically decreasing function of $t$ (as we have already pointed out). 

The shear, which is given by (see for instance \cite{Joe:2014tca}) 
\begin{equation}
\label{shear-r}
\sigma^{2}=\frac{1}{2}\sigma_{ab}\sigma^{ab}  =\frac{1}{3}\left(  \frac{\dot{p}_{b}}{p_{b}}-\frac{\dot
{p}_{c}}{p_{c}}\right)^2 ,
\end{equation}
reads explicitly as follows
\begin{equation}
\label{class-shear}
\sigma^{2}=\frac{\alpha_{0}^{2}}{3}\left(\frac{b^{2}+\gamma^{2}}{b}\right)^{2}\left(3b^{2}+\gamma^{2}\right)^{2}.
\end{equation}
From Eqs.(\ref{class-exp-scalar}), (\ref{class-theta-dot}) and (\ref{class-shear}), it is a simple exercise to see that 
$\dot{\theta}=-(1/3)\theta^{2}-2\sigma^{2}$, which is nothing but Raychaudhuri's equation. (Recall that the congruence is hypersurface orthogonal, so that there is no rotational term. In addition, the term $R_{ab}\xi^{a}\xi^{b}=R_{00}$ is identically zero on shell). 

Let us remark that by considering $K_{c}<0$, one gets that $c<0$ (since $p_{c}$ is strictly positive) and that $b>0$ (since $b$ and $c$ must have opposite signs). Exactly as above, it is shown that $p_{b}\leq \vert K_{c} \vert$. The expressions for $p_{c}$, $c$ and $p_{b}$ [respectively, Eqs. (\ref{pcb}), (\ref{cb}) and (\ref{pbb})] will be the same ones, though now with $K_{c}<0$ and $b>0$. Of course, equation (\ref{b-dot-sol}) governing the dynamics of $b$ is the same one, so is its solution (\ref{b-sol}); but now with $g>0$, provided that $b>0$. Since $\dot{b}<0$, then $b$ is a monotonically decreasing function of $t$, which implies that $g$ decreases in time. Rather than technical, the important difference between conventions $K_{c}>0$ and $K_{c}<0$ is conceptual. Recall that associated to the sign of $K_{c}$ is a time reversal, so it is  natural to write the solution to Eq.(\ref{b-dot-sol}) as
\begin{equation}
\label{b-sol-neg-k}
g(b)=2\gamma^{3}\alpha_{0}(t_{0}-t)+g(b_{0}),\qquad g(s)=\frac{\gamma s}{(s^{2}+\gamma^{2})}+\arctan\left(\frac{s}{\gamma}\right),
\end{equation}
with $t_{0}>t$ (for instance, $t_0$ would denote the `present time', whereas $t$ stands for an earlier time). Clearly, $0<g<\pi/2$ for all $b\in \mathbb{R}^{+}$, and it is a monotonically decreasing function of $t$. Equation (\ref{b-sol-neg-k}) implies that $b(t)$ will tend to $b\to \infty$ as $t\to 0$; so that, $p_{c}\to 0$, $c\to -\infty$ and $p_{b}\to 0$ as $t$ approaches zero. In particular, we have that as $t\to 0$, the Kretschmann scalar will diverge as $b^{8}$, whereas the cell volume will collapse to zero as $V\sim 1/b^{3}$ .

The explicit expression for the expansion scalar of a congruence of timelike geodesics constructed from comoving observers (i.e., with associated vector field $\xi^{a}=(\partial /\partial t )^{a}$) is also given by Eq.(\ref{class-exp-scalar}). Note that $\theta>0$ for $0<t<t_{*}$, where $t_{*}$ is `the cosmological time' at which $b(t_{*})=\gamma/\sqrt{3}$, $\theta$ is zero at $t_{*}$, and it is strictly negative for $t>t_{*}$.  The expansion scalar is, in fact, a monotonically decreasing function in time. Note, in addition, that $\theta\to \infty$ at $t=0$. The shear and the time derivative of $\theta$ have, of course, exactly the same expressions as above. 

Now, by considering the congruence of `past-directed comoving world lines', which associated vector field is $\xi^{a}=-(\partial /\partial t )^{a}$, one gets that the backward in time (BT) expansion scalar is given by
\begin{equation}
\label{expansion}
\theta_{\rm{BT}}(\tau')=-\alpha_{0}\left(\frac{b^{2}+\gamma^{2}}{b}\right)\left(3b^{2}-\gamma^{2}\right),
\end{equation}
where $b$ is evaluated at $(t_{0}-\tau')$ and the parameter $\tau'$ is from zero to $t_{0}$ (so that $t=0$ corresponds to the limit $\tau'\to t_{0}$). Thus, $\theta_{\rm{BT}}(\tau')\to -\infty$ within a finite `proper time' $t_{0}$; that is to say, the volume shrinks to zero as we approach the `initial singularity'. 

\section{Quantum Schwarzschild interior}

\label{QKS}

In this section we implement a path integral quantization of the Schwarzschild black hole interior. To do so we make use of  its Kantowski-Sachs form as well as the similarity of the latter with Bianchi I model, both being anisotropic homogeneous models. In particular, rather than starting from scratch with the LQC techniques (see e.g. \cite{Ashtekar:2011ni, Ashtekar:2009vc,Ashtekar:2005qt}) we perform a sequence of transformations in phase space that ultimately allow us to identify adequate holonomy type variables for the KS model at the hamiltonian level, first, and, second, to introduce its path integral quantization. We follow closely the analyisis for Bianchi I in  \cite{Liu:2012xp}. In this way an effective action, and hence an effective hamiltonian, can be identified from the transition amplitude of the quantum KS model. This effective hamiltonian will be used in the next
section to analyze the effective geometry for Schwarzschild interior. 

Let us notice that an effective Hamiltonian was proposed in \cite{Bohmer:2007wi,Chiou:2008eg} motivated by several previous results. Essentially it was defined by  the heuristic replacements  $b\rightarrow \sin(\mu_bb)/\mu_b$ and $c\rightarrow \sin(\mu_cc)/\mu_c$ in (\ref{class-H}), where the use  of $\mu_b,\mu_c$ follows from their appearence  in a form of the hamiltonian constraint in which curvature terms are expressed by holonomies along elementary squares of length related to them  \cite{Ashtekar:2011ni}. Our approach, that was described above, is different from this simple replacement, however, we will regain the effective hamiltonian of \cite{Chiou:2008eg}. Now various criteria turn out to be necessary in the holonomy version of the construction  \cite{Ashtekar:2011ni}. They include (i) the area of the elementary squares used
in the holonomies should not be less than the minimum area gap
$\Delta$ found in the spectrum of the area operator of the full theory, (ii) physical quantities must be independent of a  fiducial 
metric introduced along the analysis, as well as  (iii) avoidance of large quantum gravity effects in
classical regimes \cite{Ashtekar:2006uz}.  Such criteria led to propose the following form of the $\mu^{\prime}s$  \cite{Bohmer:2007wi, Chiou:2008eg, Ashtekar:2009vc}
\footnote{The two other possibilities that were explored, do not satisfy all
these criteria, yielding inconsistent physics. See for example
\cite{Joe:2014tca} and references therein.}:%
\begin{equation}
\mu_{b}:=\sqrt{\frac{\Delta}{p_{c}}}\quad\text{and}\quad\mu_{c}:=\frac
{\sqrt{\Delta p_{c}}}{p_{b}}.\label{mu}%
\end{equation}

Let us begin with the hamiltonian description and consider the classical constraint (\ref{class-H}). It is convenient to define first the following set of canonical variables \cite{Liu:2012xp}
\begin{equation} 
\lambda_{b}:=\frac{p_{b}}{\sqrt{G\hbar}},\quad\lambda_{c}:=\sqrt{\frac{{p_{c}}%
}{{G\hbar}}},\label{muprimvar}\\
\varphi_{b}:=\frac{b}{\sqrt{G\hbar}},\quad\varphi_{c}:=c\sqrt{\frac{{p_{c}}%
}{{G\hbar}}},
\end{equation}
which Poisson brackets take the form
\begin{equation} \label{PBphilambda}
\hbar\{\varphi_{l},\lambda_{j}\}=\gamma\delta_{l,j}\qquad l,j=b,c.
\end{equation}
Let us observe that the following variables 
\begin{equation}
k_{b}:=\frac{\varphi_{b}}{\lambda_{c}}=\frac{b\mu_{b}}{\sqrt{\Delta}},\quad
k_{c}:=\frac{\varphi_{c}}{\lambda_{b}}=\frac{c\mu_{c}}{\sqrt{\Delta}%
},\label{muprimvar2}%
\end{equation}
have exponential forms  
\begin{eqnarray} \label{U}
U_b := \text{ e}^{\text{i}\sqrt{\Delta}k_{b}} = \text{ e}^{\text{i}b\mu_b}, \quad
U_c :=\text{ e}^{\text{i}\sqrt{\Delta}k_{c}} = \text{ e}^{\text{i}c\mu_c},
\end{eqnarray}
that are amenable for the application of LQC techniques.
In the sequel the following identity will be of much help
\begin{equation}
\frac{\sin(\sqrt{\Delta}k_{l})}{\sqrt{\Delta}}=\frac{U_{l}-(U_{l})^{\ast}
}{2\text{i}\sqrt{\Delta}},  \quad l=b,c.\label{representacion seno}%
\end{equation}
Using the set of variables $\lambda_b,\lambda_c, k_b,k_c$, given in eqs. (\ref{mu}), (\ref{muprimvar}) and (\ref{muprimvar2}), in
the classical constraint  (\ref{class-H}) yields the following form 
\begin{equation} \label{Hmulk}
H_{\mu}=-\frac{\hbar}{2\gamma^{2}}\left[ 2(\lambda_bk_b)(\lambda_ck_c)+\frac{\lambda_{c}}{\lambda_{b}}(\lambda_bk_b)^{2}+\frac{\gamma^{2}}{G\hbar}\frac{\lambda_{b}}{\lambda_{c}}\right].
\end{equation}
Next we consider a small argument approximation for the  $(\lambda_bk_b)$ and $(\lambda_c k_c)$  factors in (\ref{Hmulk}) through
the following relations 
\begin{eqnarray}
(\lambda_b k_b) &\approx& \frac{1}{\sqrt{\Delta}} \left[ \lambda_b\sin(\sqrt{\Delta}k_b)  \right] :=\Phi_b, \label{Phib}\\
(\lambda_c k_c) &\approx& \frac{1}{\sqrt{\Delta}} \left[ \lambda_c\sin(\sqrt{\Delta}k_c)  \right] :=\Phi_c, \label{Phic}
\end{eqnarray}
where the $\sin(\sqrt{\Delta}k_b)$ and $\sin(\sqrt{\Delta}k_c)$ will be understood according to (\ref{representacion seno}) so that our 
elementary variables will be $\lambda_l,U_l, l=b,c$. Their Poisson brackets can be obtained upon combination of (\ref{PBphilambda}),
(\ref{muprimvar2}) and (\ref{U}). The result is
\begin{eqnarray} \label{PBlambdaU}
\left\{  {\lambda}_b, {U}_b\right\}&=&\frac{\ell_0}{i\hbar}   \frac{{U}_b}{{\lambda}_c}, \nonumber\\
\left\{  {\lambda}_c, {U}_c\right\} &=& \frac{\ell_0}{i\hbar}  \frac{{U}_c}{{\lambda}_b} ,
\end{eqnarray}
with $\ell_0:=\gamma\sqrt{\Delta}$.

Let us now proceed to quantization. Our elementary quantum observables will be $\widehat{\lambda}_{b},\widehat
{\lambda}_{c},\hat{U_b}$ and $\hat{U_c}$.  The
Hilbert space of this system will be $\mathcal{H}^{(2)}_{\mathrm Poly}= \mathcal{H}_{\mathrm Poly}\otimes\mathcal{H}_{\mathrm Poly}$ with 
$\mathcal{H}_{\mathrm Poly}=L^{2}(\mathbb{R}
_{\text{Bohr}},\text{d}\mu_{H})$ and  $\mathbb{R}_{\text{Bohr}}$ is the
Bohr compactification of the real line and $\text{d}\mu_{H}$ is its Haar's measure
\cite{Velhinho:2007gg}. We use a basis of eigenkets $|\vec{\lambda}\rangle:=|\lambda_{b},\lambda
_{c}\rangle$ of the operators $\hat{\lambda}_{b}$ and $\hat{\lambda}_{c}$.
These basis satisfy%
\begin{equation}
\langle\vec{\lambda}^{\prime}|\vec{\lambda}\rangle=\delta_{\vec{\lambda
}^{\prime},\vec{\lambda}},
\end{equation}
where $\delta_{\lambda^{\prime},\lambda}$ is a Kronecker delta. To represent $\hat{U}'s$ we make use of the commutation relations
\begin{eqnarray}
\left[  \hat{\lambda}_b, \hat{U}_b\right] &=& \ell_0  \frac{\hat{U}_b}{\hat{\lambda}_c}, \\
\left[  \hat{\lambda}_c, \hat{U}_c\right] &=& \ell_0  \frac{\hat{U}_c}{\hat{\lambda}_b} ,
\end{eqnarray}
which follow from the application of Dirac's prescription to the Poisson brackets (\ref{PBlambdaU}).
They lead to
\begin{equation}
\hat{U}_{b}|\vec{\lambda}\rangle=|\lambda_{b}+\ell_{0}/\lambda_{c},\lambda
_{c}\rangle,\\
\qquad\hat{U}_{c}|\vec{\lambda}\rangle=|\lambda_{b},\lambda_{c}+\ell
_{0}/\lambda_{b}\rangle.
\end{equation}
Now we proceed to implement the quantum version of (\ref{Phib}) and (\ref{Phic}) as \cite{Ashtekar:2010gz}
\begin{equation}
\hat{\Phi}_{b}:=\frac{1}{\sqrt{\Delta}}\left[  \sqrt{\hat{\lambda}_{b}}%
\widehat{\sin{\sqrt{\Delta}k_{b}}}\sqrt{\hat{\lambda}_{b}}\right]
\quad\text{and}\quad\hat{\Phi}_{c}:=\frac{1}{\sqrt{\Delta}}\left[  \sqrt
{\hat{\lambda}_{c}}\widehat{\sin{\sqrt{\Delta}k_{c}}}\sqrt{\hat{\lambda}_{c}%
}\right]  ,\label{phi}%
\end{equation}
whose action on the basis are%
\begin{align}
\hat{\Phi}_{b}|\vec{\lambda}\rangle &  =-\frac{\text{i}\sqrt{\lambda_{b}}%
}{2\sqrt{\Delta}}\left[  \sqrt{\lambda_{b}+\ell_{0}/\lambda_{c}}|\lambda
_{b}+\ell_{0}/\lambda_{c},\lambda_{c}\rangle-\sqrt{\lambda_{b}-\ell
_{0}/\lambda_{c}}|\lambda_{b}-\ell_{0}/\lambda_{c},\lambda_{c}\rangle\right]
,\label{phiactua}\\
\hat{\Phi}_{c}|\vec{\lambda}\rangle &  =-\frac{\text{i}\sqrt{\lambda_{c}}%
}{2\sqrt{\Delta}}\left[  \sqrt{\lambda_{c}+\ell_{0}/\lambda_{b}}|\lambda
_{b},\lambda_{c}+\ell_{0}/\lambda_{b}\rangle-\sqrt{\lambda_{c}-\ell
_{0}/\lambda_{b}}|\lambda_{b},\lambda_{c}-\ell_{0}/\lambda_{b}\rangle\right] 
. \label{phiactua2}
\end{align}
Hence the quantum version of the hamiltonian constraint  (\ref{Hmulk}), using (\ref{Phib}) and (\ref{Phic}) first at the classical level and then
 their quantum version (\ref{phi}), becomes
\begin{equation}
\hat{H}_{\mu}=-\frac{\hbar}{2\gamma^{2}}\left[  \hat{\Phi}_{b}\hat{\Phi}%
_{c}+\hat{\Phi}_{c}\hat{\Phi}_{b}+(\hat{\Phi}_{b})^{2}\frac{\hat{\lambda}_{c}%
}{2\hat{\lambda}_{b}}+\frac{\hat{\lambda}_{c}}{2\hat{\lambda}_{b}}\hat{(\Phi
}_{b})^{2}+\frac{\gamma^{2}}{G\hbar}\frac{\hat{\lambda}_{b}}{\hat{\lambda_{c}%
}}\right] , \label{Hmuprima}%
\end{equation}
for which a symmetric ordering has been introduced. Its building blocks are defined to act upon the eigenbasis 
$|\vec\lambda\rangle$, according to (\ref{phiactua}) and (\ref{phiactua2}) while the $\hat{\lambda}$ factors act diagonally.

This hamiltonian will be used now to obtain Feynman's formula for the propagator to go from state $|\vec{\lambda}_{i};\tau_{i}\rangle$ at proper time $\tau_i$ to $|\vec{\lambda}_{f};\tau_{f}\rangle$ at time $\tau_f > \tau_i$. It takes the form
\begin{equation}
\langle
\vec{\lambda}_{f};\tau_{f} |\vec{\lambda}_{i};\tau_{i}\rangle=\langle
\vec{\lambda}_{f}|\text{ e}^{-\text{i}\Delta\tau\hat{H}_{\mu}/\hbar}%
|\vec{\lambda}_{i}\rangle, \qquad \Delta \tau= \tau_f-\tau_i. \label{aprima1}%
\end{equation}

To calculate explicitly such propagator we consider as usual a partition of the time interval $\Delta \tau$ and split, accordingly, the time evolution operator as
\begin{equation} \label{Nexp}
\text{ e}^{-\text{i}\Delta\tau\hat{H}_{\mu}/\hbar}=\prod_{n=0}^{N-1}\text{
e}^{-\text{i}\epsilon\hat{H}_{\mu}/\hbar},\quad\text{where}\quad
N\epsilon=\Delta\tau.
\end{equation}
Then using
\begin{equation}
\hat{\mathbb{I}}=\sum_{\vec{\lambda}_{n}}|\vec{\lambda}_{n}\rangle\langle
\vec{\lambda}_{n}|,
\end{equation}
together with  (\ref{Nexp}) allow us to rewrite  (\ref{aprima1}) as%
\begin{equation}
\langle \vec{\lambda}_{f};\tau_{f} |\vec{\lambda}_{i};\tau_{i}\rangle=\sum
_{\vec{\lambda}_{N-1},...,\vec{\lambda}_{1}}\prod_{n=0}^{N-1}\langle
\vec{\lambda}_{n+1}|\text{e}^{-\text{i}\epsilon\hat{H}_{\mu}/\hbar}%
|\vec{\lambda}_{n}\rangle, \label{aprima}%
\end{equation}
where $\vec{\lambda}_{f}=\vec{\lambda}_{N}$ y $\vec{\lambda}_{i}=\vec{\lambda
}_{0}$. Next we consider that for small $\epsilon$%
\begin{equation}
\langle\vec{\lambda}_{n+1}|\text{e}^{-\text{i}\epsilon\hat{H}_{\mu}/\hbar
}|\vec{\lambda}_{n}\rangle=\delta_{\vec{\lambda}_{n+1},\vec{\lambda}_{n}%
}-\text{i}\frac{\epsilon}{\hbar}\langle\vec{\lambda}_{n+1}|\hat{H}_{\mu}%
|\vec{\lambda}_{n}\rangle+\mathcal{O}(\epsilon^{2}). \label{aprimepartial}%
\end{equation}
The matrix elements $\langle\vec{\lambda}_{n+1}|\hat{H}_{\mu}|\vec{\lambda
}_{n}\rangle$ can be calculated using (\ref{phiactua}) and (\ref{Hmuprima}):
\begin{align} \label{LHL}
\langle\vec{\lambda}_{n+1}|\hat{H}_{\mu}|\vec{\lambda}_{n}\rangle
=\frac{\hbar}{8\gamma^{2}\Delta}&\left\{\sqrt{\lambda_{b,n+1}\lambda_{b,n}}\sqrt{\lambda_{c,n}\lambda_{c,n+1}}\left(P_n +Q_n\right)+\frac{4\gamma^{2}\Delta}{G\hbar}\frac{\lambda_{b,n}}{\lambda
_{c,n}}\delta_{\vec{\lambda}_{n}\vec{\lambda}_{n+1}}\right. \nonumber\\
&\hspace{1cm} \left.  +\sqrt{\lambda_{b,n}\lambda_{b,n+1}}\frac{\lambda_{b,n}+\lambda_{b,n+1}}{2}\left(  \frac{\lambda_{c,n}}{2\lambda_{b,n}}+\frac{\lambda_{c,n+1}}{2\lambda_{b,n+1}}\right)R_n \right\}.
\end{align}
where
\begin{align} \label{LHL}
P_n&=\left(  \delta_{\lambda_{b,n+1},\lambda_{b,n}+\ell_{0}/\lambda_{c,n+1}}-\delta_{\lambda_{b,n+1},\lambda_{b,n}-\ell_{0}
/\lambda_{c,n+1}}\right) \left(  \delta_{\lambda_{c,n+1},\lambda_{c,n}+\ell_{0}/\lambda_{b,n}}-\delta_{\lambda_{c,n+1},\lambda_{c,n}-\ell
_{0}/\lambda_{b,n}}\right),\nonumber\\
Q_n&=\left(  \delta_{\lambda_{b,n+1},\lambda_{b,n}+\ell_{0}/\lambda_{c,n}}-\delta_{\lambda_{b,n+1},\lambda_{b,n}-\ell_{0}/\lambda_{c,n} }\right)  \left(  \delta_{\lambda_{c,n+1},\lambda_{c,n}+\ell_{0}/\lambda_{b,n+1}}-\delta_{\lambda_{c,n+1},\lambda_{c,n}-\ell_{0}%
/\lambda_{b,n+1}}\right), \nonumber\\
R_n&=\left(  \delta_{\lambda_{b,n+1},\lambda_{b,n}+2\ell_{0}/\lambda_{c,n+1}}-2\delta_{\lambda_{b,n+1},\lambda_{b,n}}+\delta
_{\lambda_{b,n+1},\lambda_{b,n}-2\ell_{0}/\lambda_{c,n+1}}\right)\delta_{\lambda_{c,n+1},\lambda_{c,n}}.
\end{align}

At this point we can see from (\ref{phiactua})-(\ref{phiactua2}) and hence in the matrix elements of $\hat{H}_{\mu}$  given by Eq. (\ref{Hmuprima}), that states supported on a regular  (equally spaced) $\vec{\lambda}-$lattice do not fit into our quantum KS model.
This is a difficulty that also appears in the Bianchi I models and thus, to proceed further,  we can use the approximation proposed in \cite{Liu:2012xp} for that case. It
consists of exploiting the fact that we are looking for a continuous yet quantum effective approximation \cite{Varadarajan:1999it,Ashtekar:2002vh}. Hence, effectively, one can  replace at leading order the Kronecker deltas by Dirac's in (\ref{LHL}). This implies
that one is approximating at leading order a description from ${\cal H}_{\mathrm Poly}^{(2)}$ to ${\cal H}_{\mathrm Sch}^{(2)}={\cal H}_{\mathrm Sch}\otimes {\cal H}_{\mathrm Sch}, {\cal H}_{\mathrm Sch}=L^2(\mathbb{R},dx)$, so that $\vec{\lambda}$ is now a
continuous variable. Within this approximation it is useful to adopt the following integral form of Dirac's delta 
\begin{equation}
\delta(\lambda_{n+1}-\lambda_{n})=\frac{1}{2\pi\gamma}\int_{\mathbb{R}%
}\text{d}\varphi_{n+1}\text{ e}^{-\text{i}\varphi_{n+1}(\lambda_{n+1}%
-\lambda_{n})/\gamma}.
\end{equation}
Then, eq. (\ref{aprimepartial}) can be expressed as
\begin{align}
\langle\vec{\lambda}_{n+1}|\text{e}^{-\text{i}\epsilon\hat{H}_{\mu}}%
|\vec{\lambda}_{n}\rangle=&\left(  \frac{1}{2\pi\gamma}\right)  ^{2}\int
d\vec{\varphi}_{n+1}\text{ e}^{-\text{i}\vec{\varphi}_{n+1}(\vec{\lambda}_{n+1}%
-\vec{\lambda}_{n})/\gamma}\nonumber\\ 
&\hspace{1.5cm}\times\left\{  1+\text{i}\frac{\epsilon}{2\gamma
^{2}\Delta}\left[M_n+N_n +  L_n+\frac{\gamma^{2}\Delta}{G\hbar}%
\frac{\lambda_{b,n}}{\lambda_{c,n}}\right]  \right\}  +\mathcal{O}%
(\epsilon^{2}),
\end{align}
where
\begin{align}
M_n&=\sqrt{\lambda_{b,n+1}\lambda_{b,n}}\sqrt{\lambda_{c,n}\lambda_{c,n+1}}\sin(\sqrt{\Delta}\varphi_{b,n+1}/\lambda_{c,n+1})\sin(\sqrt{\Delta}\varphi_{c,n+1}/\lambda_{b,n}),\\
N_n&=\sqrt{\lambda_{b,n+1}\lambda_{b,n}}\sqrt{\lambda_{c,n}\lambda_{c,n+1}}\sin(\sqrt{\Delta}\varphi_{b,n+1}/\lambda_{c,n})\sin(\sqrt{\Delta}\varphi_{c,n+1}/\lambda_{b,n+1}),\\
L_n&=\sqrt{\lambda_{b,n}\lambda_{b,n+1}}\frac{\lambda_{b,n}+\lambda_{b,n+1}}{2}\left(  \frac{\lambda_{c,n}}{2\lambda_{b,n}}+\frac{\lambda_{c,n+1}}{2\lambda_{b,n+1}}\right)\sin(\sqrt{\Delta}\varphi_{b,n+1}/\lambda_{c,n+1})^{2}.
\end{align}
here $\vec{\varphi}=(\varphi_{b},\varphi_{c})$. This last expression allow us
to rewrite the propagator in the form%
\begin{equation} \label{TA}
\langle \vec{\lambda}_{f};\tau_{f} |\vec{\lambda}_{i};\tau_{i}\rangle =\left(
\frac{1}{2\pi\gamma}\right)  ^{2N} \int d{\vec{\lambda}_{N-1}... d\vec{\lambda
}_{1}}\int\text{d}\vec{\varphi}_{N}...\text{d}\vec{\varphi}_{1}\text{ e}^{\text{i}/\hbar
S_{\mu}^{N}}+\mathcal{O}(\epsilon^{2}),
\end{equation}
where%
\begin{align}\label{SNprime}
S_{\mu}^{N}  &  =\epsilon\sum_{n=0}^{N-1}-\frac{\hbar}{\gamma}\vec{\varphi
}_{n+1}\frac{\vec{\lambda}_{n+1}-\vec{\lambda}_{n}}{\epsilon}+\frac{\hbar
}{2\gamma^{2}\Delta}\left[  \sqrt{\lambda_{b,n+1}\lambda_{b,n}}\sqrt
{\lambda_{c,n}\lambda_{c,n+1}}\right. \nonumber \\
&  \times\left(  \sin(\sqrt{\Delta}\varphi_{b,n+1}/\lambda_{c,n+1})\sin
(\sqrt{\Delta}\varphi_{c,n+1}/\lambda_{b,n})+\sin(\sqrt{\Delta}\varphi
_{b,n+1}/\lambda_{c,n})\sin(\sqrt{\Delta}\varphi_{c,n+1}/\lambda
_{b,n+1})\right) \nonumber\\
&  \left.  +\sqrt{\lambda_{b,n}\lambda_{b,n+1}}\frac{\lambda_{b,n}%
+\lambda_{b,n+1}}{2}\left(  \frac{\lambda_{c,n}}{2\lambda_{b,n}}+\frac
{\lambda_{c,n+1}}{2\lambda_{b,n+1}}\right)  \sin(\sqrt{\Delta}\varphi
_{b,n+1}/\lambda_{c,n+1})^{2}+\frac{\gamma^{2}\Delta}{G\hbar}\frac
{\lambda_{b,n}}{\lambda_{c,n}}\right]  .
\end{align}
Now we take the limit $N\rightarrow\infty$ and  Eq.  (\ref{SNprime}) takes the form%
\begin{align}
S_{\mu}=  &  \lim_{N\rightarrow\infty}S_{\mu}^{N}= \int_{\tau_{i}}^{\tau_{f}%
}\text{d}\tau \left\{ -\frac{\hbar}{\gamma}\vec{\varphi}\cdot\dot{\vec{\lambda}%
}\nonumber\label{Sprime} \right.\\
&  \left. +\frac{\hbar}{2\gamma^{2}\Delta}\left[  \lambda_{b}\lambda_{c}\sin
(\sqrt{\Delta}\varphi_{b}/\lambda_{c})\left(  2\sin(\sqrt{\Delta}\varphi
_{c}/\lambda_{b})+\sin(\sqrt{\Delta}\varphi_{b}/\lambda_{c})\right)
+\frac{\gamma^{2}\Delta}{G\hbar}\frac{\lambda_{b}}{\lambda_{c}}\right]  \right\}.
\end{align}
Therefore we can see that the effective hamiltonian is
\begin{equation}
H_{\mu}^{\text{eff}}=-\frac{\hbar}{2\gamma^{2}\Delta}\left[  \lambda
_{b}\lambda_{c}\sin(\sqrt{\Delta}\varphi_{b}/\lambda_{c})\left(  2\sin
(\sqrt{\Delta}\varphi_{c}/\lambda_{b})+\sin(\sqrt{\Delta}\varphi_{b}%
/\lambda_{c})\right)  +\frac{\gamma^{2}\Delta}{G\hbar}\frac{\lambda^{b}%
}{\lambda_{c}}\right]  .
\end{equation}
Using (\ref{muprimvar}) to return to the original variables $(b,c,p_{b}%
,p_{c})$ we get finally
\begin{equation}
H_{\mu}^{\text{eff}}=-\frac{1}{2G\gamma^{2}}\left[  2\sqrt{p_{c}}\frac{\sin
\mu_{b}b}{\mu_{b}}\frac{\sin\mu_{c}c}{\mu_{c}}+\frac{p_{b}}{\sqrt{p_{c}}%
}\left(  \frac{\sin\mu_{b}b}{\mu_{b}}\right)  ^{2}+\gamma^{2}\frac{p_{b}%
}{\sqrt{p_{c}}}\right]  . \label{eff ham}%
\end{equation}

Let us emphasize that the classical hamiltonian (\ref{class-H}) is recovered by taking the small argument limit $\vert \mu_{l}l\vert <<1$ ($l=b,c$) in the effective hamiltonian (\ref{eff ham}); i.e., the classical model, namely, the classical hamiltonian and equations of motion, are recovered from the effective one in the regime $\vert \mu_{l}l\vert <<1$.

The hamiltonian (\ref{eff ham}) is the key piece defining and governing the effective quantum geometry. Effective states $(b,c,p_{b},p_{c})$ lie in the constraint surface $H_{\mu}^{\text{eff}}=0$, ``evolving" along the gauge integral curves of the hamiltonian vector field generated by $H_{\mu}^{\text{eff}}$. In the next section we will focus on analyze how geometrical quantities behave in the effective quantum scenario provided 
by $H_{\mu}^{\text{eff}}$.

\section{Effective loop quantum dynamics}

\label{EffectiveKS}

To investigate the effective geometry, let us begin by considering the Hamilton equations associated to the hamiltonian (\ref{eff ham}), $\dot{\zeta}=\{\zeta,H_{\mu}^{\text{eff}}\}$, with the Poisson brackets  (\ref{poisson-brackets}), and the effective scalar constraint $H_{\mu}^{\text{eff}}=0$,
\begin{equation}
\dot{b}=\frac{-\gamma^{2}\mu_{b}^{2}-\sin\left(b\mu_{b}\right)
\left[\sin\left(b\mu_{b}\right) -2c\mu_{c}\cos\left(  c\mu_{c}\right)   +2\sin\left(  c\mu_{c}\right)  \right]  }{2\gamma
\sqrt{\Delta}\mu_{b}}, \label{beffdot}%
\end{equation}

\begin{equation}
\dot{c}=\frac{\gamma^{2}\mu_{b}^{2}+2b\mu_{b}\cos\left(  b\mu
_{b}\right)  \left[  \sin\left(  b\mu_{b}\right)  +\sin\left(
c\mu_{c}\right)  \right]  -\sin\left(  b\mu_{b}\right)  \left[
\sin\left(  b\mu_{b}\right)+2c\mu_{c}\cos\left(  c\mu_{c}\right)   +2\sin\left(  c\mu_{c}\right)  \right]  }{2\gamma
\sqrt{\Delta}\mu_{c}}, \label{ceffdot}%
\end{equation}

\begin{equation}
\dot{p}_{b}=\frac{\sqrt{\Delta}\cos\left(  b\mu_{b}\right)  \left[
\sin\left(  b\mu_{b}\right)  +\sin\left(  c\mu_{c}\right)
\right]  }{\gamma\mu_{b}\mu_{c}}, \label{pbeffdot}%
\end{equation}

\begin{equation}
\dot{p}_{c}=\frac{2\sqrt{\Delta}\cos\left(  c\mu_{c}\right)
\sin\left(  b\mu_{b}\right)  }{\gamma\mu_{b}^{2}}. \label{pceffdot}%
\end{equation}

At this point two remarks are in order. First, provided that both $p_{b}$ and $p_{c}$ are strictly positive quantities, the first term in (\ref{eff ham}) must be strictly negative in order to satisfy the constraint. Second, since $\gamma \mu_{b}\mu_{c}=\gamma \Delta /p_{b}$ and $\gamma \mu_{b}^{2}=\gamma \Delta /p_{c}$ [cf. Eq.  (\ref{mu})], Eq.(\ref{pbeffdot}) and Eq.(\ref{pceffdot}) can be written in the form $d(\ln p_{i})/dt=f_{i}(b\mu_{b},c\mu_{c})$, $i=a,b$.

A solution to the effective model is a sufficiently smooth{\footnote{Let $f=(l,p_{l})$, $l=a,b$, be a solution to Eqs. (\ref{beffdot})-(\ref{pceffdot}), and let us suppose that $f$ is, at least, of class $C^{1}$. Thus, it follows from Eqs. (\ref{beffdot})-(\ref{pceffdot}) that $f$ is, in fact, a $C^{\infty}$ function.}}, real  solution to Eqs. (\ref{beffdot})-(\ref{pceffdot}) which, in addition, satisfies the scalar constraint (\ref{eff ham}). Let us refer to solutions of the effective model as effective solutions. Since the dynamics is pure gauge, each point in the constraint surface is an appropriate initial condition for effective solutions. Now, to fix notation, let $\chi_{0}$ be the initial condition $(b_{0},c_{0},p_{b0},p_{c0})$ at $t=t_0$ (the reference initial time) to the effective solution $\chi=(b,c,p_{b},p_{c})$.

From Eq.  (\ref{eff ham}), it follows that effective solutions $\chi$ must satisfy, in particular, that
\begin{equation}
\sin(\mu_{b}b)=-\sin(\mu_{c}c)\pm \sqrt{\sin^{2}(\mu_{c}c)-\frac{\Delta \gamma^2}{p_{c}}}.
\end{equation}
Since effective solutions $\chi$ are real ones, the discriminant must necessarily be nonnegative, so that $\sin^{2}(\mu_{c}c)\geq \Delta \gamma^2/p_{c}$. Thus, in particular, we have that $p_{c}$ is bounded from below\footnote{This bound is consistent with that found in
Ref. \cite{Chiou:2008nm}, where $p_{c}\geq\Delta\gamma^{2}/3.$} by $\Delta\gamma^{2}$; i.e.,  
\begin{equation}
p_{c}\geq\Delta\gamma^{2}. \label{cota pc}%
\end{equation}
This expression implies that the area of $S^{2}$ [cf. Eq. (\ref{leng-are-vol})] cannot be less than $4\pi\Delta\gamma^{2}$ in the effective geometry. 
 By using inequality (\ref{cota pc}) into the relations defining $\mu_{b}$ and $\mu_{c}$ [cf. Eq. (\ref{mu})], we get that
\begin{equation}
\label{mub-muc-fbounds}
 \frac{\Delta \gamma}{p_{c}}\leq \mu_{b}\leq \frac{1}{\gamma},\qquad 
\frac{\Delta \gamma}{p_{b}}\leq \mu_{c}\leq \frac{p_{c}}{\gamma p_{b}}.
\end{equation}
Thus, in particular, 
\begin{equation}
\label{cota mub}
\gamma\mu_{b}\leq1. 
\end{equation}

Using again that the first term in (\ref{eff ham}) must be strictly negative it must be the case that $\sin(\mu_{b}b)$ and $\sin(\mu_{c}c)$ must have opposite constant signs,
\begin{equation}
\sin(\mu_{l}l)>0,\quad \sin(\mu_{l'}l')<0, \label{sinmubsinmuc}%
\end{equation}
with $l$ being equal to $b$ or $c$, and $l'$ being the complementary of $l$; that is, for $l=b$ ($l=c$), $l'=c$ ($l'=b$). Strict inequalities (\ref{sinmubsinmuc}), and the continuity of the functions $\mu_{b}b$ and $\mu_{c}c$, imply that $2n_{0}\pi<\mu_{l}l <(2n_{0}+1)\pi$ and $(2m_{0}-1)\pi<\mu_{l'}l' <2m_{0}\pi$, for some $n_{0},m_{0}\in \mathbb{Z}$ fixed and determined by the effective solution $\chi$; in fact, by its corresponding initial condition $\chi_0$. Indeed, given an initial condition $\chi_{0}$, we will have that $\sin(\mu_{l}l)_{0}>0$ and that $\sin(\mu_{l'}l')_{0}<0$, so that $n_{0}$ is the greatest integer $n$ satisfying that $n<(\mu_{l}l)_{0}/2\pi$, whereas $m_{0}$ is the least integer $m$ satisfying that $(\mu_{l'}l')_{0}/2\pi<m$. By continuity, $\mu_{l}l$ and $\mu_{l'}l'$ must remain, respectively, in $\big{(}2n_{0}\pi,(2n_{0}+1)\pi\big{)}$ and in $\big{(}(2m_{0}-1)\pi,2m_{0}\pi\big{)}$; otherwise, the constraint will be violated. We then have disjoint sectors, and they are as many as the distinct pairs $(n_{0},m_{0})$ that the initial conditions define. Although we will perform our analysis by considering a generic sector, it is worth remarking that it is only within the $(0,0)$-sector that the regime $\mu_{d} \vert d\vert <<1$ can be consistently treated.

Let us introduce a more symmetric notation through
$$ N_{l}= 
\begin{cases}
  (2n_{0}+1)  ,& \text{if } n_{0}\geq 0\\
    2\vert n_{0}\vert ,              & \text{if }  n_{0}<0
\end{cases}, \qquad
N_{l'}= 
\begin{cases}
  2m_{0}  ,& \text{if } m_{0}\geq 1\\
    (2\vert m_{0}\vert +1) ,              & \text{if }  m_{0}\leq 0
\end{cases} $$
In terms of $N_{d}$ (with $d$ being $b$ or $c$) we have that $\mu_{d}\vert d \vert$ is confined to be in $\big( \pi(N_{d}-1),\pi N_{d}\big)$. Explicitly,  given an effective solution $\chi$, the quantities $\mu_{b}\vert b \vert$ and $\mu_{c}\vert c \vert$ are bounded by $\pi(N_{b}-1)<\mu_{b}\vert b \vert <\pi N_{b}$ and by $ \pi(N_{c}-1)<\mu_{c}\vert c \vert <\pi N_{c}$, where $N_{b}$ and $N_{c}$ are (strictly) positive fixed integers determined by the initial condition $\chi_{0}$.

Now, inequalities $\pi(N_{d}-1)<\mu_{d}\vert d \vert$ and $\mu_{d}\vert d \vert <\pi N_{d}$ imply that $0<\vert\sin(\mu_{d}d)\vert \leq 1$ and that $0\leq \vert\cos(\mu_{d}d)\vert < 1$. Thus, for any given phase-space function $g$, we will have the strict inequality
\begin{equation}
\label{cos-stbound}
 \vert g\cos(\mu_{d}d)\vert< \vert g \vert. 
\end{equation}
In particular, we have that the strict inequality $\vert\sin(\mu_{d}d)\cos(\mu_{d'}d')\vert<1$ must be satisfied. 

Now, let us consider $\vert \dot{p}_{b}\vert$ and $\vert \dot{p}_{c}\vert$. From Eqs. (\ref{pbeffdot})-(\ref{pceffdot}) -written in terms of the explicit expressions for $\gamma \mu_{b}\mu_{c}$ and $\gamma \mu_{b}^{2}$- it follows by using the triangle inequality, the boundedness of the sine function and Eq.(\ref{cos-stbound}) that 
\begin{equation}
\label{dpb-dpc-bounds}
\left\vert \dot{p}_{b}\right\vert < \left(\frac{3}{2\gamma \sqrt{\Delta}}\right)p_{b}, \qquad \left\vert \dot{p}_{c}\right\vert < \left(\frac{2}{\gamma \sqrt{\Delta}}\right)p_{c}.
\end{equation}
To get the first inequality we have used, in addition, the relationship $\sin(2b\mu_{b})=2\cos(b\mu_{b})\sin(b\mu_{b})$ in Eq. (\ref{pbeffdot}). Similar calculations employing relations (\ref{cota pc})-(\ref{mub-muc-fbounds}) in Eqs. (\ref{beffdot}) and (\ref{ceffdot})  shows that  $\vert \dot{b}\vert$ and $\vert \dot{c}\vert$ are bounded from above by
\begin{equation}
\vert \dot{b}\vert < \frac{1}{2\sqrt{\Delta}}+\left(\frac{3+2\mu_{c}\vert c \vert}{2\gamma^{2}\Delta^{3/2}}\right)p_{c}\leq \left(\frac{2+\mu_{c}\vert c\vert}{\gamma^{2}\Delta^{3/2}}\right)p_{c},\quad \vert \dot{c}\vert <\left(\frac{4+2\mu_{c}\vert c \vert+3\mu_{b}\vert b \vert}{2\gamma^{2}\Delta^{3/2}}\right) p_{b}.
\end{equation}
Since $\mu_{d}\vert d\vert< \pi N_{d}$, we obtain that
\begin{equation}
\label{db-dc-bounds}
\vert \dot{b}\vert <  \left(\frac{2+\pi N_{c}}{\gamma^{2}\Delta^{3/2}}\right)p_{c}
,\quad \vert \dot{c}\vert < \left(\frac{4+ 2\pi N_{c}+3\pi N_{b}} {2\gamma^{2}\Delta^{3/2}}\right) p_{b}.
\end{equation}

 Inequalities (\ref{dpb-dpc-bounds}) and (\ref{db-dc-bounds}) imply that $\vert \dot{\chi} \vert$ is bounded from above by $F_{(N_{b},N_{c})}\vert \chi \vert$, where
$$F_{(N_{b},N_{c})}^{2}=\max\left\{\frac{9}{4\gamma^{2}\Delta}+\frac{(4+2\pi N_{c}+3\pi N_{b})^{2}}{4\gamma^{4}\Delta^{3}}\, , \,\frac{4}{\gamma^{2}\Delta}+\frac{(2+\pi N_{c})^{2}}{\gamma^{4}\Delta^{3}}\right\}.$$ 
All effective solutions  in the sector labelled by $(N_{b},N_{c})$  turn out  defined for $t\in \mathbb{R}$. 
In addition, let us remark that effective solutions are bounded by the exponential function. Indeed, recall that Eq.(\ref{pbeffdot}) and Eq.(\ref{pceffdot}) can be written in the form $d(\ln p_{d})/dt=f_{d}(b\mu_{b},c\mu_{c})$. Thus, combining Eqs. (\ref{pceffdot}) and (\ref{cota pc}), employing the boundedness of the sine and Eq. (\ref{cos-stbound}), it is not difficult to see that
\begin{equation}
\label{pc-bounds}
\Delta\gamma^{2}\, \leq \, p_{c} \, <\, p_{c0}\:e^{ 2\vert t-t_{0} \vert/ \gamma{\sqrt{\Delta}}}.
\end{equation}
Similarly, from Eq.(\ref{pbeffdot}) it follows that
\begin{equation}
\label{pb-bounds}
p_{b0}\:e^{-3\vert t-t_{0}\vert/\gamma\sqrt{4\Delta}}< p_{b}<p_{b0}\:e^{3\vert t-t_{0}\vert/\gamma\sqrt{4\Delta}}.
\end{equation}
Since $\pi(N_{d}-1)<\mu_{d}\vert d \vert <\pi N_{d}$, using inequalities (\ref{cota pc}) and (\ref{pc-bounds})-(\ref{pb-bounds}), we get that
\begin{equation}
\label{b-mod-bound}
\gamma \pi (N_{b}-1)<\vert b \vert < \frac{\pi N_{b}}{(\mu_{b})_{0}}e^{\vert t-t_{0} \vert/ \gamma{\sqrt{\Delta}}},
\end{equation}
\begin{equation}
\label{c-mod-bound}
\frac{\pi (N_{c}-1)}{(\mu_{c})_{0}}e^{-5\vert t-t_{0}\vert/\gamma\sqrt{4\Delta}}<\vert c \vert < \frac{\pi N_{c}\,p_{b0}}{\Delta \gamma}e^{3\vert t-t_{0}\vert/\gamma\sqrt{4\Delta}}
\end{equation}
Thus, in contrast to the classical model, $b$ and $c$ are finite quantities at every time: there is not a finite proper time limit, $t_f$, at which $b$ and $c$ will become infinite. Besides, {\emph{for all}} $t\in \mathbb{R}$, $p_{c}$ is bounded from below by a positive number, namely $\Delta \gamma^{2}$, and $p_{b}$ is a strictly positive quantity as well. Note, in addition, that Eqs. (\ref{pc-bounds})-(\ref{pb-bounds}) prevent the metric (\ref{metrica cinematica}) to have a coordinate singularity in the effective approach.

Let us now focus on the behavior of geometrical and invariant quantities. From Eqs. (\ref{pbeffdot})-(\ref{pceffdot}), using the explicit expressions for $\gamma \mu_{b}\mu_{c}$ and $\gamma \mu_{b}^{2}$, it immediately follows that 
\begin{equation}
\label{dtheta-b}
\left\vert \theta_{\rm{eff}}\right\vert =\left\vert\frac{\dot{p}_{b}}{p_{b}}+\frac{\dot{p}_{c}}{2p_{c}}\right\vert = \frac{1}{\gamma\sqrt{\Delta}} \left\vert \sin(b\mu_{b}+c\mu_{c})+\frac{1}{2}\sin(2b\mu_{b})\right\vert \leq  \frac{3}{\gamma\sqrt{4\Delta}}, 
\end{equation}
\begin{equation}
\label{dshear-b}
\left\vert \sigma_{\rm{eff}}\right\vert  = \frac{1}{\sqrt{3}}\left\vert\frac{\dot{p}_{b}}{p_{b}}-\frac{\dot{p}_{c}}{p_{c}}\right\vert 
=\frac{1}{\gamma\sqrt{3\Delta}} \left\vert  \sin(c\mu_{c}-b\mu_{b})-\cos(c\mu_{c})\sin(b\mu_{b})+\frac{1}{2}\sin(2b\mu_{b})\ \right\vert
<  \frac{5}{\gamma\sqrt{12\Delta}},
\end{equation}
where we have used Eq.(\ref{cos-stbound}) to get the strict inequality in the last term of Eq. (\ref{dshear-b}). The boundedness of the expansion scalar ensures, in particular, that the volume of a cell will remain different from zero at any finite proper time. Indeed, let $V_{r}=(4\pi p_{b}\sqrt{p_{c}})\,\vert_{t_{r}}$ be the volume of the cell ${\cal{V}}=[0,L]\times S^{2}$ at an arbitrary reference proper time $t_{r}$ of comoving observers in the effective Kantowski-Sachs geometry (with $N=1$), and let $t$ be any other finite proper time. It is a simple matter to see that $\vert \theta_{\rm{eff}}\vert\leq 3/\gamma\sqrt{4\Delta}$ implies that 
\begin{equation}
\label{two-bounds}
V_{r}e^{-3\vert t-t_{r}\vert/\gamma\sqrt{4\Delta}}\leq V\leq V_{r}e^{3\vert t-t_{r}\vert/\gamma\sqrt{4\Delta}},
\end{equation}
where $V$ is the volume of the cell $\cal{V}$ at time $t$. The volume $V$ is a well-defined, strictly positive quantity at any finite proper time $t\in \mathbb{R}$ and, consequently, the congruence of timelike geodesics defined by comoving observers [i.e., the integral curves of the vector field $\xi^{a}=(\partial/\partial t)^{a}$] will not develop a caustic (at finite proper times).

Let us now demonstrate that the effective Ricci and Kretschmann scalars, $R_{\rm{eff}}$ and $K_{\rm{eff}}$, are in fact well-behaved, finite quantities. Provided that $R_{\rm{eff}}$ and $K_{\rm{eff}}$ have second order terms in the time derivatives of $p_{b}$ and $p_{c}$, we shall first calculate $\{\dot{p}_{d},H_{\mu}^{\rm{eff}}\}$. By using Eqs. (\ref{pbeffdot})-(\ref{pceffdot}), as well as Eqs. (\ref{beffdot})-(\ref{ceffdot}), a straightforward calculation shows that 
\begin{eqnarray}
\label{pbeffdotdot}
\ddot{p}_{b}= \frac{p_{b}}{2\gamma^{2}\Delta}\Bigg{(}
&\gamma^{2}&\mu_{b}^{2}\sin(2b\mu_{b})\left[\sin(b\mu_{b})\cos(c\mu_{c})+\frac{1}{4}\sin(2c\mu_{c})\right]\nonumber \\
\, &+&\sin(2c\mu_{c})\left[\sin(c\mu_{c})\cos(b\mu_{b})+\frac{1}{4}\sin(2b\mu_{b})\sin^{2}(b\mu_{b})\right]\nonumber \\
\, &+& \cos(c\mu_{c})\left[\sin(2c\mu_{c})\cos(b\mu_{b})+\cos(c\mu_{c})\sin^{2}(b\mu_{b})\sin(2b\mu_{b})\right](\mu_{b}b-\mu_{c}c)\Bigg{)},
\end{eqnarray}

\begin{eqnarray}
\label{pceffdotdot}
\ddot{p}_{c}=-\cos(b\mu_{b}&-&c\mu_{c})+\frac{p_{c}}{\gamma^{2}\Delta}\Bigg{(}2\sin^{2}(b\mu_{b})\left[1+\cos^{2}(c\mu_{c})\right]-\frac{1}{2}\sin(2b\mu_{b})\sin(2c\mu_{c})\nonumber \\ \, &-& \sin^{2}(b\mu_{b})\cos(b\mu_{b}+c\mu_{c})
+\sin(2b\mu_{b})\left[1+\sin(b\mu_{b})\sin(c\mu_{c})\right](\mu_{c}c-\mu_{b}b)\Bigg{)}.
\end{eqnarray}

Employing the triangle inequality, condition (\ref{cos-stbound}), and the boundedness of $\mu_{b}$ [cf. (\ref{cota mub})] as well as of the sine function, we get that
\begin{equation}
\left\vert\ddot{p}_{b}\right\vert< \frac{1}{\gamma^{2}\Delta}\left(\frac{5}{4}+\big{[}\mu_{b}\vert b\vert +\mu_{c}\vert c\vert\big{]}\right)p_{b},\quad \left\vert\ddot{p}_{c}\right\vert < 1+\frac{1}{\gamma^{2}\Delta}\left(\frac{11}{2}+2\big{[}\mu_{b}\vert b\vert +\mu_{c}\vert c\vert\big{]}\right)p_{c}.
\end{equation}
Using that $\mu_{d}\vert d \vert < \pi N_{d}$, we arrive to
\begin{equation}
\label{pb-pb-ddot-bounds}
\left\vert\ddot{p}_{b}\right\vert<  \left(\frac{5+{4\pi [N_{b}+ N_{c}]}}{4\gamma^{2}\Delta}\right)p_{b},
\quad \left\vert\ddot{p}_{c}\right\vert< 
1+\left(\frac{11+{4\pi [N_{b}+ N_{c}]}}{2\gamma^{2}\Delta}\right)p_{c}\leq \left(\frac{13+{4\pi [N_{b}+ N_{c}]}}{2\gamma^{2}\Delta}\right)p_{c},
\end{equation}
where the last inequality in the second expression follows from $1\leq p_{c}/(\gamma^{2}\Delta)$. 

In order to simplify notation, let us introduce the quotients $x:=\dot{p_{b}}/p_{b}$, $y:=\dot{p_{c}}/p_{c}$, $v:=\ddot{p_{b}}/p_{b}$ and $w:=\ddot{p_{c}}/p_{c}$. So, inequalities (\ref{dpb-dpc-bounds}) and (\ref{pb-pb-ddot-bounds}) read as follows
\begin{equation}
\label{quotients-bounds}
\vert x \vert < \frac{3}{2\gamma\sqrt{\Delta}},\quad \vert y \vert < \frac{2}{\gamma\sqrt{\Delta}}, \quad \vert v \vert <
 \left(\frac{5+{4\pi [N_{b}+ N_{c}]}}{4\gamma^{2}\Delta}\right)
,\quad \vert w \vert < \left(\frac{13+{4\pi [N_{b}+ N_{c}]}}{2\gamma^{2}\Delta}\right).
\end{equation}

The Ricci scalar, which is given by $R_{\rm{eff}}=2v+w+(2/p_{c})$ [cf. Eq. (\ref{ricciscalarkinematic})], is thus bounded by
\begin{equation}
\vert R_{\rm{eff}}\vert \leq 2 \vert v \vert + \vert w \vert+\frac{2}{p_{c}}. 
\end{equation}
By using Eq. (\ref{cota pc}) and Eq. (\ref{quotients-bounds}) we have that the Ricci scalar [in the sector labelled by $(N_{b},N_{c})$] is bounded by
\begin{equation} \label{Rbound}
\vert R_{\rm{eff}}\vert <\left(\frac{11+{4\pi [N_{b}+ N_{c}]}}{\gamma^{2}\Delta}\right).
\end{equation} 

Let us now focus on the Kretschmann scalar, $K_{\rm{eff}}$. From Eq. (\ref{Kretschmannscalarkinematic}), it is easy to see that in terms of the quotients $x$, $y$, $v$ and $w$, $K_{\rm{eff}}$ is given by 
\begin{eqnarray}
K_{\rm{eff}}=&4&v^{2}+3w^{2}-4vw-8vxy+4wyx+6vy^2\nonumber \\
\,&-&5wy^{2} +6x^{2}y^{2}-8xy^{3}+\frac{7}{2}y^{4}+\frac{2}{p_{c}}y^{2}+\frac{4}{p_{c}^{2}}.
\end{eqnarray}
Clearly, the effective Kretschmann scalar turns out to be a bounded quantity. The explicit bound is obtained by using the inequalities (\ref{cota pc}) and (\ref{quotients-bounds}), as well as the triangle inequality. A straightforward calculation shows that
\begin{equation} \label{Kbound}
\vert K_{\rm{eff}} \vert < \frac{\xi}{\gamma^{4}\Delta^{2}}, \qquad \xi=4\left(6[N_{b}+N_{c}]^{2}\pi^{2}+59[N_{b}+N_{c}] \pi+160\right)+\frac{23}{2}.\end{equation}

In addition, since $\dot{\theta}_{\rm{eff}}=v-x^{2}+(w-y^{2})/2$, we get for effective solutions that
\begin{equation}
\left \vert \dot{\theta}_{{\rm{eff}}}\right\vert\leq \vert v \vert+x^{2}+\frac{1}{2}\vert w \vert +\frac{1}{2}y^{2}<\frac{1}{\gamma^{2}\Delta}\left(\frac{35}{4}+ 2(N_{b}+N_{c})\pi \right)
\end{equation}
This, together with Eqs. (\ref{dtheta-b})-(\ref{dshear-b}), proves that $(R_{00})_{\rm{eff}}$ is a bounded quantity as well.

In general, we have that any quantity of the form
\begin{equation}
\label{g-form}
\Lambda:=\sum_{j=1}^{N}C_{j}\left(\ddot{p}_{b}\right)^{n_{j}}\left(\ddot{p}_{c}\right)^{m_{j}}\left(\dot{p}_{b}\right)^{r_{j}}\left(\dot{p}_{c}\right)^{s_{j}}\left(p_{b}\right)^{\alpha_{j}}\left(p_{c}\right)^{\beta_{j}},
\end{equation}
where $n_{j}$, $m_{j}$, $r_{j}$ and $s_{j}$ are nonnegative integers, and $\alpha_{j}$ and $\beta_{j}$ are any two real numbers, is a bounded quantity on shell. Indeed, from inequalities (\ref{dpb-dpc-bounds}) and (\ref{pb-pb-ddot-bounds}) it follows that
\begin{equation}
\label{general-b}
\vert \Lambda_{\rm{eff}} \vert < \sum_{j=1}^{N}\vert C_{j}\vert\left(\frac{A_{bc}}{4\gamma^{2}\Delta}\right)^{n_{j}}\left(\frac{B_{bc}}{2\gamma^{2}\Delta}\right)^{m_{j}}
\left(\frac{3}{2\gamma\sqrt{\Delta}}\right)^{r_{j}} \left(\frac{2}{\gamma\sqrt{\Delta}}\right)^{s_{j}}
\left(p_{b}\right)^{n_{j}+r_{j}+\alpha_{j}}\left(p_{c}\right)^{m_{j}+s_{j}+\beta_{j}},
\end{equation}
where $A_{bc}:=5+{4\pi [N_{b}+ N_{c}]}$ and  $B_{bc}:=13+{4\pi [N_{b}+ N_{c}]}$. Then, we have that in the effective approach of the KS model, any effective quantity $\Lambda_{\rm{eff}}$ of the form (\ref{g-form}) will be bounded by (\ref{general-b}). Provided that any scalar polynomial invariant $P$ associated to the metric (\ref{metrica cinematica}) -with the lapse function being set to the unit constant function- will take the form (\ref{g-form}), as it is actually the case for the Ricci and Kretschmann scalars, we can assert that in the effective geometry of the KS model $P_{\rm{eff}}$ will be bounded everywhere, even though its classical counterpart is not (i.e., even if $P_{\rm{class}}$ diverges at some regime).

\section{Discussion}
\label{sec-disc}
The quest for the fundamental nature of spacetime may shed light on long standing problems as the singularities appearing in classical general relativity and the ultraviolet divergences of field theories. Hence quantum gravity theories that endow with quantum character to spacetime acquire particular interest. Loop quantum gravity, in particular, has yielded homogenous cosmological models in which the classical singularity is replaced by a quantum bounce and thus Schwarzschild interior which classically amounts to a homogeneous, Kantowski-Sachs, model is amenable for a similar treatment. Indeed the loop quantization of Schwarschild interior showed that the would be classical singularity is actually traversable and later on some heuristic effective models confirmed the same result but also added possible replacements for the singularities like another black hole, a Nariai universe or a white hole. However, connecting the quantum treatment with the effective model was left open. In this paper we have advanced a proposal that links the loop quantum description of the Schwarzschild interior with an effective model that is based on a path integral scheme. Specifically we have built a transition amplitude between two loop quantum states of the Kantowski-Sacks model as a path integral, Eq. (\ref{aprima}), consisting of an imaginary exponential of an action in phase space from which the effective heuristic hamiltonian constraint descends, Eqs. (\ref{TA}), (\ref{Sprime}) and (\ref{eff ham}). Although this strategy was originally used for homogeneous isotropic, as well as some anisotropic, models the particular case of Kantowski-Sachs had not been dealt with before. 

Armed with the effective constraint we embarked in the study of the ensuing dynamics that happened to lead to rather simple analytic bounds for the basic phase space variables, Eqs. (\ref{pc-bounds})-(\ref{c-mod-bound}) and their time derivatives. In particular expansion and shear turn out to be bounded too as it is the volume, Eqs. (\ref{dtheta-b})-(\ref{two-bounds}). Similarly, by considering the second order time derivatives of the basic phase space variables, according to the effective dynamics, we get that both the effective Ricci  and Kretschmann scalars, Eqs. (\ref{Rbound}) and (\ref{Kbound}), are bounded. This bounded character actually holds for any product  of the form (\ref{g-form}) containing second order and first order time derivatives as well as powers of the variables $p_b,p_c$. It is a remarkable fact that analytic results were obtained from the effective dynamics which, although simple in appearance, could only be treated numerically in previous works.

There are several interesting points which can be further explored along the lines we have followed in the present work. One of them concerns 
our analysis performed considering a generic sector labeled by $(N_l,N_{l'})$ indicated by the detailed form of the effective hamiltonian constraint. Since it is in the regime $\mu_{d} \vert d\vert <<1$, from which the classical behavior can be recovered through a semiclassical approximation this selects only the $(0,0)$-sector. Thus, it is natural to ask about the physical relevance of the other sectors. Another thing we have not done in the present work is an analysis of the would be classical horizon. Since we have adopted the improved quantization for the Kantowski-Sachs model it is expected that it will differ from the recent results of \cite{Corichi:2015xia} that adopts an effective dynamics preserving the classical horizon definition. Further work is required to clarify other possible physical differences. 
Indeed, for example, recent phenomenological results on black hole evaporation \cite{Barrau:2016qri,Barrau:2015ana} require connecting interior effective descriptions like the one studied presently with that corresponding to the exterior. It would be interesting to combine our path integral analysis of the Schwarszchild interior including a coupling to a scalar field along the lines of \cite{Hartle:1976tp} and extend it to the exterior region in order to investigate further quantum gravity corrections to the black hole emission (See e.g. \cite{Tecotl:2015cya} which applies polymer path integral to a mechanical model to study the problematics of the black hole semiclassical approximation.)
Finally, important consequences of the features we have found here may play a role in the geodesic analysis in regard to completeness and perhaps complement recent results for the cosmological case \cite{Joe:2014tca,Saini:2016vgo}.

\section*{Acknowledgements}
This work was partially supported by CONACyT Grant No. 237351 ``Implicaciones f\'isicas de la estructura del espacio tiempo" and DGAPA-UNAM Grant No. IN113115 ``Teor\'ia de campos en fondos curvos, gravedad cu\'antica y holograf\'ia". 

\appendix{}%

\end{document}